\documentclass{article}
\usepackage{arxiv, color, soul}

\usepackage{amsmath, amsthm, mathabx, bbm}
\usepackage{authblk}
\usepackage{mathtools}
\usepackage{txfonts}

\usepackage[utf8]{inputenc} % allow utf-8 input
\usepackage[T1]{fontenc}    % use 8-bit T1 fonts
\usepackage{hyperref}       % hyperlinks
\usepackage{url}            % simple URL typesetting
\usepackage{booktabs}       % professional-quality tables
\usepackage{amsfonts}       % blackboard math symbols
\usepackage{nicefrac}       % compact symbols for 1/2, etc.
\usepackage{microtype}      % microtypography
\usepackage{lipsum}
\usepackage{natbib}
\usepackage{diagbox}
\usepackage{bm}
\usepackage{caption}

\theoremstyle{definition}% default

\DeclareMathOperator*{\argmin}{arg\,min}
\DeclareMathOperator{\tr}{tr}
\DeclareMathOperator{\diag}{diag}
\DeclareMathOperator{\Herm}{Herm}
\DeclareMathOperator{\Span}{span}
\DeclareMathOperator{\cut}{cut}
\DeclareMathOperator{\conv}{conv}

\DeclareMathOperator{\Int}{int}
\DeclareMathOperator{\UBD}{UBD}

\title{Natural evolution strategies and { Variational Monte Carlo}}

\author[1]{Tianchen Zhao}
\author[2]{Giuseppe Carleo}
\author[3,4]{James Stokes}
\author[1,3]{Shravan Veerapaneni}

\affil[1]{Department of Mathematics, University of Michigan, Ann Arbor, MI 48109, USA}
\affil[2]{Institute of Physics, EPFL, CH-1015 Lausanne, Switzerland}% \authorcr\{\tt glenn, joe\}@nvidia.com}
\affil[3]{Center for Computational Mathematics, Flatiron Institute, New York, NY 10010, USA}
\affil[4]{Center for Computational Quantum Physics, Flatiron Institute, New York, NY 10010, USA}
\begin{document}
\maketitle

\begin{abstract}
A notion of quantum natural evolution strategies is introduced, which provides a geometric synthesis of a number of known quantum/classical algorithms for performing classical black-box optimization. Recent work of \cite{gomes2019classical} on {heuristic} combinatorial optimization using neural quantum states is pedagogically reviewed in this context, emphasizing the connection with natural evolution strategies. The algorithmic framework is illustrated for approximate combinatorial optimization problems, and a systematic strategy is found for improving the approximation ratios. In particular it is found that natural evolution strategies can achieve approximation ratios {competitive with widely used heuristic algorithms} for Max-Cut, at the expense of increased computation time.

\end{abstract}

\section{Introduction}
An evolution strategy \citep{schwefel1977numerische, rechenberg1978evolutionsstrategien} is a black-box optimization algorithm that iteratively updates a population of candidates within the feasible region of the search space. The population is updated by a process of random mutation, followed by fitness evaluation, and subsequent recombination of best-performing members to form the next generation. The focus of this paper is on the natural evolution strategies (NES) algorithm \citep{wierstra2014natural} and its quantum variants, in which the population is represented by a smoothly parameterized family of search distributions defined on the search space. The mutation is achieved by sampling new candidates from the search distribution, which yields a gradient estimator of the expected fitness. The recombination step involves updating the parameters of the search distribution in the direction of steepest ascent, with respect to the information geometry implicit in the choice of search distribution.

Natural evolution strategies has recently demonstrated considerable progress in solving black-box optimization problems in high dimensions, including continuous optimization problems relevant to reinforcement learning \citep{salimans2017evolution}. Comparatively little work has been done on the discrete optimization side (see however \citep{,malago2008information, malago2011towards, ollivier2017information}).
It was very recently shown by \cite{gomes2019classical} that techniques from quantum variational Monte Carlo literature \citep{carleo2017solving} can be adapted for {approximate heuristic solution of}  combinatorial optimization problems. Meanwhile, in the quantum computation literature, considerable effort has focused on {approximate heuristic solution of}  combinatorial optimization problems using low-depth quantum circuits \citep{farhi2014quantum, yao2020policy}.

The unifying principle shared by the above algorithms is their utilization of Monte-Carlo samples drawn from a particular probability distribution, the choice of which is determined by a local optimization problem over a variational family of search distributions. The algorithms differ in the choice of variational family, as well as the geometry underlying the local optimization problem, and the use of quantum or classical resources to perform sampling.  In this work, we provide a unified view on the relationship between the geometries which underpin these algorithms. In addition, we revisit the experiments of \cite{gomes2019classical} { in which heuristic algorithms were found to outperform NES both in terms of approximation ratio and time to solution. In contrast, we show that} NES can achieve approximation ratios for Max-Cut { competitive with widely used algorithms}, albeit at the expense of significantly increased computational cost. The key factor impacting performance is identified to be the batch size of the stochastic gradient estimator.

The paper is structured as follows. First, the natural evolution strategies algorithm is reviewed, highlighting its information-geometric origin. Next, we clarify the relationship between natural evolution strategies and quantum approximate optimization including the classical/quantum hybrid approach introduced in \citep{gomes2019classical}. Finally,  numerical experiments are presented, focusing on the problem of combinatorial optimization for the Max-Cut problem.

\section{Background}\label{sec:background}
Consider the problem of optimizing a real-valued, but otherwise arbitrary function $f \in \mathbb{R}^X = \{{X} \to \mathbb{R}\}$ defined on a search space ${X}$. For simplicity of presentation we focus on $|{X}| < \infty$, although the method generalizes in a straightforward way to infinite search spaces, as required for continuous optimization. Now consider the probability simplex $\mathcal{P}({X})$ defined as,
\begin{equation}
    \mathcal{P}(X) = \Big\{ p \in \mathbb{R}^X : p \succeq 0, \: \sum_{x\in X} p(x) = 1 \Big\} \enspace .
\end{equation}
The natural evolution strategies algorithm can be motivated by the following two observations, which concern the convex and Riemannian geometry of the set $\mathcal{P}(X)$: 

\textbf{1.} The optimization problem admits the following equivalent convex relaxation,
\begin{equation}\label{e:cvx}
    \min_{x \in {X}} f(x) = \min_{p \in \mathcal{P}({X})}\Big( \underset{x\sim p}{\mathbb{E}} [f(x)] \Big) \enspace ,
\end{equation}
whose global minimizers form the subsimplex consisting of the following convex hull of Dirac distributions,
\begin{equation}
    \conv \Big\{\delta_{a} \in \mathcal{P}(X) :a \in \argmin_{x \in {X}} f(x) \Big\} \enspace .
\end{equation}

\textbf{2.} There is a natural Riemannian metric called the Fisher-Rao metric defined on the interior $\Int(\mathcal{P}({X}))$ of the probability simplex, consisting of strictly positive probability vectors. The distance between $p,q \succ 0$ is defined as
\begin{equation}\label{e:dFR}
    d_{\rm FR}(p,q) := \arccos{(\left\langle \sqrt{p}, \sqrt{q} \right\rangle)} \enspace ,
\end{equation}
where $\sqrt{p}$ denotes the elementwise square root of the probability vector $p$.

Given a smoothly parametrized family of search distributions $\{ p_\theta :\theta \in \mathbb{R}^d\} \subseteq \mathcal{P}({X})$, one obtains a trivial variational bound as follows,
\begin{equation}
    \min_{x \in {X}} f(x) \leq \min_{\theta \in \mathbb{R}^d}\Big( \underset{x\sim p_\theta}{\mathbb{E}} [f(x)] \Big) =: \min_{\theta \in \mathbb{R}^d} L(\theta) \enspace .
\end{equation}

Natural evolution strategies seeks to obtain a good approximation ratio by optimizing the above variational upper bound using Riemannian gradient descent in the geometry induced by the Fisher-Rao metric, otherwise known as natural gradient descent \citep{amari1998natural}. 

Specifically, given a learning rate $\eta > 0$ and initial condition $\theta_0 \in \mathbb{R}^d$, one considers the deterministic sequence in $\mathbb{R}^d$ defined by,
\begin{equation}\label{e:ngd}
    \theta_{t+1} = \argmin_{\theta \in \mathbb{R}^d} \left[ \langle \theta - \theta_{t}, \nabla L(\theta_t) \rangle + \frac{1}{2\eta}\Vert \theta - \theta_t \Vert^2_{I_{\theta_t}}\right] \enspace ,
\end{equation}
where $I_{\theta}$ denotes the Fisher information matrix, evaluated at the parameter vector $\theta \in \mathbb{R}^d$,
\begin{equation}\label{e:fisher}
    I_\theta = \underset{x \sim p_\theta}{\mathbb{E}} \left[\nabla_\theta \log p_\theta(x) \otimes \nabla_\theta \log p_\theta (x) \right] \enspace .
\end{equation}
Indeed, it can be shown that $I_\theta$ is the local coordinate representation of the Riemannian metric tensor induced by \eqref{e:dFR}, and the restriction to the interior of the probability simplex ensures that the logarithm is defined\footnote{This discussion ignored the fact that, unlike a bone fide Riemannian metric, the Fisher information matrix can be degenerate, in which case we choose the minimizer of \eqref{e:ngd} to be $\theta_{t+1} = \theta_t - \eta I_{\theta_t}^{+} \nabla L(\theta_t)$, where $I_\theta^+$ denotes the pseudo-inverse of $I_\theta$.}.

Observe that the iteration \eqref{e:ngd} defining the sequence $(\theta_t)_{t \geq 0}$ involves the unknown function $f$. The natural evolution strategies can now be defined as the randomized algorithm inspired by \eqref{e:ngd}, in which $\nabla L(\theta)$ is replaced by a stochastic gradient estimator obtained by sampling from the search distribution $p_\theta$. Likewise, if the Fisher information \eqref{e:fisher} cannot be evaluated in closed form, then it can be replaced by an associated estimator.

\section{Quantum approximate optimization as natural evolution strategies}
{In this section we pedagogically review the proposal of \cite{gomes2019classical} showing it to be a variant of Natural Evolution Strategies in which the optimization dynamics is modified as a consequence of the quantum state geometry. In addition, we identify the regime in which both methods coincide.}
The quantization of natural evolution strategies  proceeds by replacing the search space $X$ by a complex Euclidean space,
\begin{equation}
    \mathbb{C}^{X} = \Span \{ |x\rangle : x \in X \} \enspace ,
\end{equation}
whose orthonormal basis elements are $|x\rangle$. Moreover, the probability simplex $\mathcal{P}(X)$ is replaced by the convex set of density operators,
\begin{equation}
    \mathcal{D}(\mathbb{C}^X) = \{\rho \in \Herm(\mathbb{C}^X) : \rho \succeq 0 , \: \tr(\rho) = 1 \} \enspace ,
\end{equation}
where $\Herm(\mathbb{C}^X)$ denotes the set of Hermitian operators on $\mathbb{C}^X$. It is clear that any classical probability distribution $p \in \mathcal{P}(X)$ can be encoded as the following diagonal density operator $\diag(p) := \sum_{x \in X} p(x) |x \rangle \langle x |$. It is equally clear that this does not extinguish the space of admissible density operators: another possibility being the rank-1  projection operator
 $P_\psi = |\psi \rangle \langle \psi |/\langle \psi | \psi \rangle$ onto the one-dimensional subspace spanned by the vector $\psi \in \mathbb{C}^X$. Any admissible density operator $\rho \in \mathcal{D}(\mathbb{C}^X)$ gives rise to a valid probability distribution, which we call $\diag(\rho) \in \mathcal{P}(X)$, obtained from the diagonal matrix representation in the standard basis.

It is conceptually useful to introduce the following Hermitian operator $H_f \in \Herm(\mathbb{C}^X)$ (diagonalized by the standard basis),
\begin{equation}\label{e:Hf}
    H_f(x) := \sum_{x \in {X}} f(x) |x \rangle \langle x | \enspace ,
\end{equation}
whose ground-state subspace encodes the solution of the optimization problem,
\begin{equation}
    \Span \Big\{ |a\rangle : a \in \argmin_{x\in X} f(x) \Big\} \enspace .
\end{equation}
Then for any density operator $\rho \in \mathcal{D}(\mathbb{C}^X)$ we see that the quantum expectation value of $H_f$ evaluated in the state $\rho$, is computed by the following classical expectation value,
\begin{equation}\label{e:qtmev}
    \tr(\rho H_f) 
    = \underset{x\sim \diag(\rho)}{\mathbb{E}}\left[ f(x)\right] \enspace \enspace ,
\end{equation}
which is an obvious upper bound for $\min_{x \in X} f(x)$.  The above identity demonstrates the widely known fact that for diagonal operators of the form \eqref{e:Hf}, unconstrained optimization over the space of quantum states $\mathcal{D}(\mathbb{C}^X)$ is equivalent to optimization over the probability simplex $\mathcal{P}(X)$.

\cite{gomes2019classical} asks if constrained optimization within a parametrized subset of density operators provides a useful heuristic for approximate combinatorial optimization. In particular, they consider the case of rank-1 projectors, for which there exists a natural Riemannian metric called the Fubini-Study metric defined as follows,
\begin{equation}
    d_{\rm FS}(P_\psi,P_\phi) := \arccos\left(\sqrt{\tr(P_\psi P_\phi)}\right) \enspace .
\end{equation}

Thus, given a smoothly parametrized subset $\{ \psi_\theta : \theta \in \mathbb{R}^d \} \subseteq \mathbb{C}^X$ of a complex Euclidean space, one can define quantum natural evolution strategies as the local optimization of the following variational upper bound, via Riemannian gradient descent in the geometry induced by the Fubini-Study metric,
\begin{equation}\label{e:fubini}
    \min_{x \in {X}} f(x) \leq 
    \min_{\theta \in \mathbb{R}^d}
    \Big(
    \underset{x \sim |\psi_\theta|^2}{\mathbb{E}} \big[f(x)\big]\Big) \enspace .
\end{equation}
The choice to restrict to rank-1 projection operators involves no loss of generality compared to classical natural evolution strategies because if we choose $\psi_\theta = \sum_{x \in {X}} \sqrt{p_\theta(x)} | x \rangle$, then the Fubini-Study geometry reduces to Fisher-Rao. Thus, if the parametric family is chosen to be strictly positive $\psi_\theta \succ 0$, then Riemannian gradient descent in the Fubini-Study geometry coincides with natural gradient descent\footnote{The local coordinate representation of the Fubini-Study and Fisher-Rao metric tensor agree if the Berry connection vanishes \citep{stokes2019quantum}.} and we recover classical natural evolution strategies. Therefore we henceforth use the terminology `natural gradient' to refer to both geometries interchangeably.

The natural gradient has been thoroughly explored in the variational Monte Carlo for ground-state optimization of non-diagonal Hermitian operators  \citep{sorella_aps98,carleo2017solving}, and more recently in the variational quantum algorithm literature \citep{yuan2019theory, stokes2019quantum, koczor2019quantum}.  Finally, we note the possibility of generalizing to higher-rank density operators, the investigation of which is left to future work.

\section{Experiments}
Consider combinatorial optimization problems defined on  $X = \{\pm 1\}^n$. For concreteness we focus on the Max-Cut problem corresponding to an undirected graph $G=(V,E)$ of size $|V|=n$. The function $f \in \mathbb{R}^X$ to be minimized is simply,
\begin{equation}
    f(x) = \sum_{\{i,j\} \in E}\frac{x_i x_j - 1}{2}  \enspace ,
\end{equation}
where $-f(x)$ is the size of the cut corresponding to the configuration $x \in X$.
After fixing a parametrized family of wavefunctions, we locally optimize the variational upper bound \eqref{e:fubini} using stochastic natural gradient descent. Following \cite{gomes2019classical}, we choose the variational wavefunction $\psi_\theta \in \mathbb{C}^X$ to be of Boltzmann form \citep{carleo2017solving},
\begin{equation}\label{e:rbm}
    \psi_\theta(x) =  \sum_{z \in \{\pm 1\}^m} \exp\big[\langle z,W x + b\rangle + \langle x, c\rangle\big] \enspace ,
\end{equation}
where the variational parameters $\theta = (W, b, c) \in \mathbb{F}^{m\times n}\times \mathbb{F}^m \times \mathbb{F}^n$ and $\mathbb{F}$ denotes either $\mathbb{R}$ or $\mathbb{C}$. In the case $\mathbb{F}=\mathbb{R}$ we have $\psi_\theta \succ 0$ and the optimization problem is equivalent to natural evolution strategies  \citep{wierstra2014natural}. An example illustrating the increased expressiveness of complex restricted Boltzmann machines compared to their real-valued counterparts for representing classical probability distributions is presented in the supplementary material.

\subsection{Performance Evaluation on Max-Cut problem}
Although no polynomial-time algorithm is known for solving Max-Cut on general graphs, many approximation algorithms have been developed in the past decades. Random Cut Algorithm is a
simple randomized 0.5-approximation algorithm that randomly assigns each node to a partition (e.g., see \cite{mitzenmacher-cup05}). \cite{goemans1995improved} improved the performance ratio from 0.5 to at least 0.87856, by making use of the semidefinite programming (SDP) relaxation of the original integer quadratic program. \cite{burer-mp01} reformulated the SDP for Max-Cut into a non-convex problem, with a benefit of having lower dimension and no conic constraint, {but with the disadvantage of being heuristic in nature.}

The above { heuristic and approximate} algorithms were used as benchmark solvers for comparison with quantum natural evolution strategies. The implementation of Goemans-Williamson Algorithm used the \texttt{CVXPY}~\citep{cvxpy,cvxpy_rewriting} package and the Burer-Monteiro reformulation with the Riemannian Trust-Region method~\citep{absil-fcm07} used \texttt{Manopt} toolbox~\citep{boumal-jml14}, which essentially implements the optimization algorithm proposed by \citep{journee-siam10}.

The classical and quantum variants of natural evolution strategies were realized using the variational Monte Carlo (VMC)~\citep{mcmillan-pr65} method with Stochastic Reconfiguration (SR) \citep{sorella_aps98}, as implemented in the \texttt{NetKet} toolbox \citep{carleo-swx19}.
The SR optimization was performed using a regularization parameter $\lambda=0.1$ and a learning rate $\eta = 5\times 10^{-2}$, for 90 iterations. At each iteration, the number of Monte Carlo sampled observables (batch size for training) is 4096. The mean performance of the observable batch drawn from the trained model is reported.
For Restricted Boltzmann Machine (RBM) model, the number of hidden variables is set to be the same as the number of spins; the weights are complex-valued, initialized with Gaussian distribution of mean 0 and standard deviation (std) 0.01.
Throughout the experiments, the timing benchmarks are performed on a core of an 8-core processor, Intel(R) Xeon(R) CPU E5-2650 v2 @ 2.60GHz, with 128 GB of memory.

\begin{figure}[h!]
\centering
\includegraphics[width=0.48\linewidth]{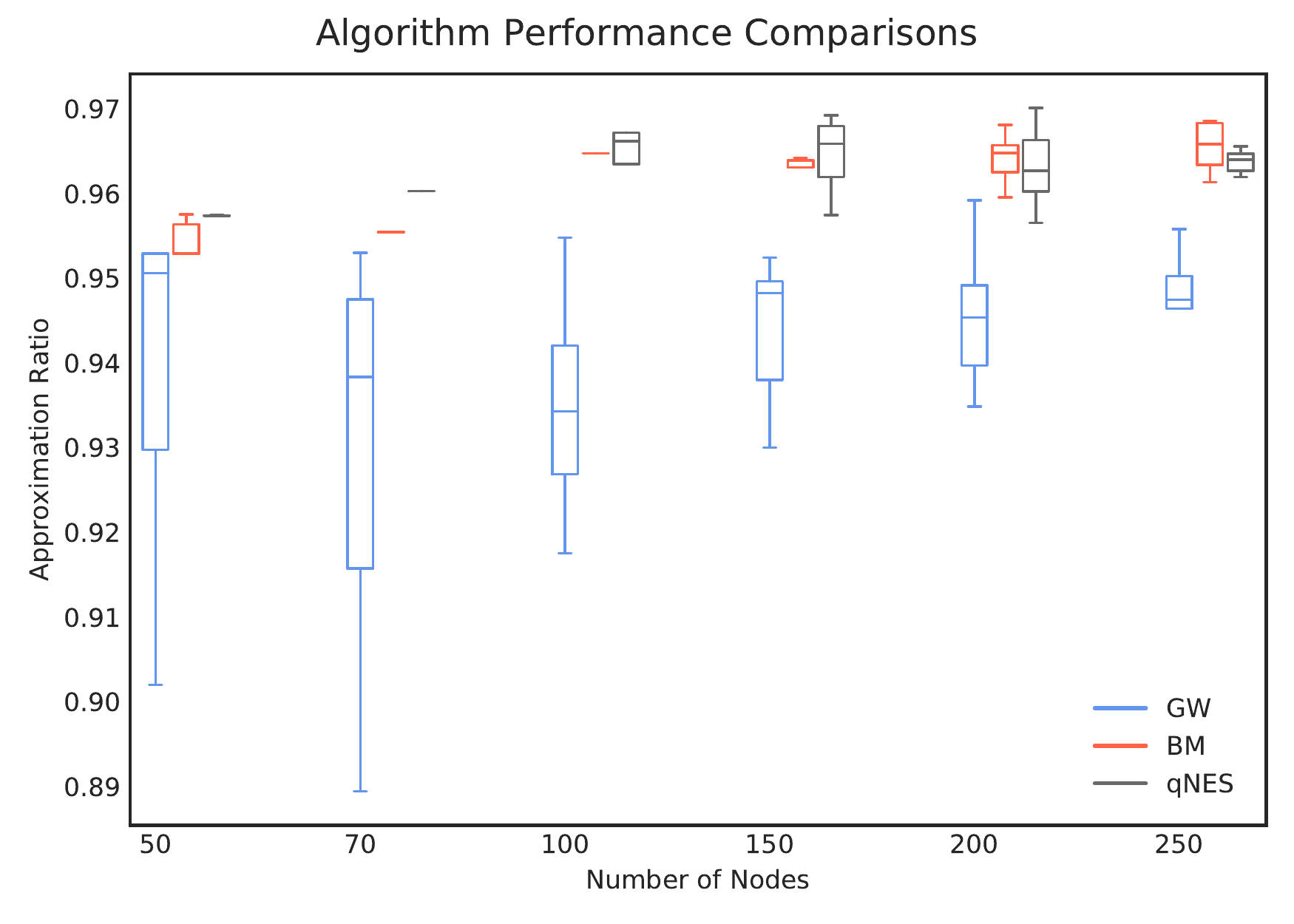}
\includegraphics[width=0.48\linewidth]{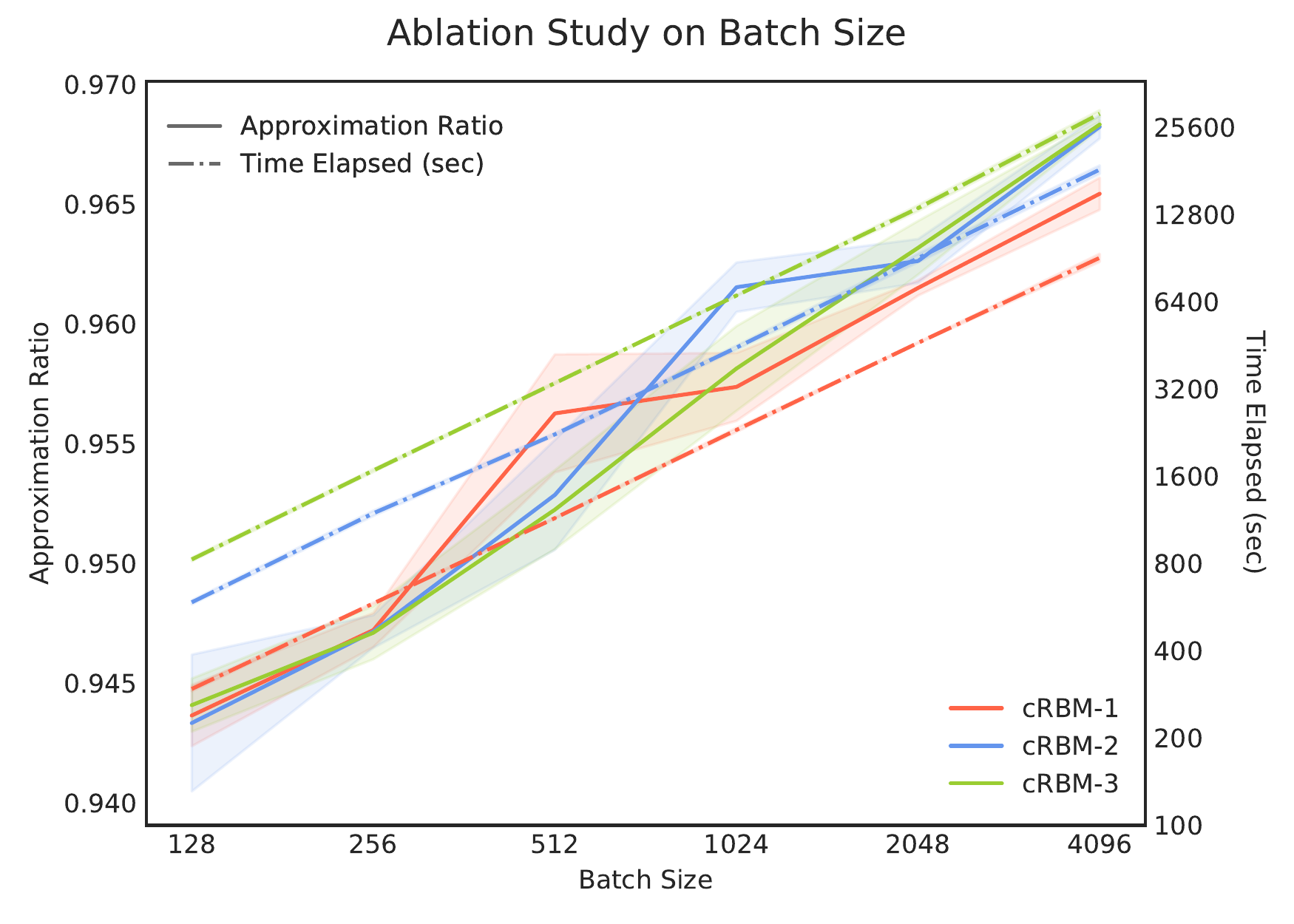}
\vspace{-2px}
\caption{\small (L) The performance comparison box plot for Goemans-Williamson (GW), Burer–Monteiro (BM) and quantum Natural Evolution Strategies (qNES). The approximation ratio is approximated by dividing the upper bound from the cut number. The performance of qNES is on par with the SDP benchmarks. (R) An ablation study for the training batch size, tested on the graph instance with 150 nodes, using cRBMs with different hidden unit density. The increase in batch size improves the performance of qNES, at a cost of the linear increase in training time.}
\label{fig:main_result_ratio}
\end{figure}

For evaluation, we constructed a problem instance for each graph size $n$ by randomly generating graph Laplacians with edge density $50\%$, for $n \in \{50, 70, 100, 150, 200, 250\}$. This defines a set of problem instances, indexed by $n$, which are held fixed throughout the experiments. For fixed problem instance, each algorithm was executed 10 times using 10 random seeds/initializations. In Table~\ref{tab:main_result_cut}, we report the mean and std of the performance over graph instances of different sizes. Since the optimal cut for a given problem instance cannot be computed for large scale problems, we approximate it with an upper bound $\UBD(I)$~\citep{boumal-neurips16}, which is the optimal value of the SDP relaxation, according to the arguments in \citep{goemans1995improved}. In Figure~\ref{fig:main_result_ratio}(L), we present the ratio $\cut(A(I))/\UBD(I)$ for the same algorithms $A$ and graph instances $I$ in Table~\ref{tab:main_result_cut} with box plot.

\begin{table*}[htbp]
\setlength{\tabcolsep}{2pt}

\caption*{\small Algorithm Performance Comparisons}

\scriptsize
\centering
\begin{tabular}[t]{c|c|c|c|c|c|c}
\hline
\diagbox[width=8.5em]{Cut Number}{\# of Nodes} & 50 & 70 & 100 & 150 & 200 & 250 \\
\hline
\hline
\multicolumn{7}{c}{Cut Number}\\
\hline
Random
& 149.60 $\pm$ 7.41 & 297.10 $\pm$ 11.48 & 614.30 $\pm$ 16.94 & 1436.80 $\pm$ 27.14 & 2467.30 $\pm$ 30.27 & 3888.00 $\pm$ 39.06 \\
\hline
Goemans-Williamson
& 203.40 $\pm$ 3.61 & 380.90 $\pm$ 8.48 & 752.50 $\pm$ 9.22 & 1685.70 $\pm$ 13.10 & 2875.10 $\pm$ 22.34 & 4439.90 $\pm$ 26.07 \\
\hline
Burer–Monteiro
& 206.30 $\pm$ 0.46 & 390.90 $\pm$ 0.54 & 776.60 $\pm$ 1.56 & 1719.90 $\pm$ 1.58 & $\bm{2931.20} \pm \bm{8.07}$ & $\bm{4526.70} \pm \bm{12.96}$ \\
\hline
Natural Evolution Strategies
& $\bm{206.97} \pm \bm{0.01}$ & $\bm{392.98} \pm \bm{0.01}$ & $\bm{777.62} \pm \bm{1.40}$ & $\bm{1721.84} \pm \bm{7.21}$ & 2927.82 $\pm$ 12.10 & 4515.72 $\pm$ 11.41 \\
\hline
\hline
\multicolumn{7}{c}{Time elapsed (sec)}\\
\hline
Random
& * & * & * & * & * & * \\
\hline
Goemans-Williamson
& 0.32 $\pm$ 0.11 & 0.51 $\pm$ 0.04 & 1.59 $\pm$ 0.02 & 5.34 $\pm$ 0.20 & 12.24 $\pm$ 0.34 & 21.09 $\pm$ 0.54 \\
\hline
Burer–Monteiro
& 0.77 $\pm$ 0.12 & 0.96 $\pm$ 0.08 & 1.36 $\pm$ 0.08 & 1.52 $\pm$ 0.16 & 2.32 $\pm$ 0.13 & 2.63 $\pm$ 0.14 \\
\hline
Natural Evolution Strategies
& 964.20 $\pm$ 12.58 & 1883.46 $\pm$ 20.39 & 3683.93 $\pm$ 65.30 & 8606.25 $\pm$ 203.14 & 15693.06 $\pm$ 431.68 & 23621.38 $\pm$ 847.84 \\
\hline
\end{tabular}
\caption{\small Performance comparison for algorithms on graph instances of different sizes. The mean and std are computed from 10 trials with different random seeds. Here, "*" indicates that the time elapsed is less than 0.1 seconds. }
\vspace{-7px}
\label{tab:main_result_cut}
\end{table*}

The choice of batch size was found to be a crucial factor controlling the performance of the algorithm. Intuitively, it quantifies the exploration capability in the state space: the algorithm has a better chance to discover the ground state if it is allowed to explore more.  The performance (as measured by approximation ratio) as well as the training time, as a function of batch size are reported in Figure~\ref{fig:main_result_ratio}(R) on the graph instance with 150 nodes. 
The ablation study also reveals that increasing hidden unit density is correlated with performance, provided that training batch size is correspondingly increased. This is expected behavior since increased model capacity is required to capture the increasing complexity from the sampled observables.

\begin{table*}[htbp!]
\setlength{\tabcolsep}{2pt}

\caption*{\small Ablation Study on Optimizer and Model Architecture}
\scriptsize
\centering
\begin{tabular}[t]{c|c|c|c|c|c}
\hline
\diagbox[width=8.5em]{Architecture}{Optimizer} & Adadelta & Adamax & Momentum & RMSprop & SGD \\
\hline
\hline
\multicolumn{6}{c}{w.o. Natural Gradient}\\
\hline
cRBM-1 & 1692.25 $\pm$ 6.13 & 1648.19 $\pm$ 12.67 & 1656.60 $\pm$ 6.83 & 1642.60 $\pm$ 15.07 & 1608.20 $\pm$ 52.05 \\
\hline
\hline
\multicolumn{6}{c}{Natural Gradient}\\
\hline
rRBM-1 & 1646.96 $\pm$ 6.80 & 1687.16 $\pm$ 10.34 & 1694.02 $\pm$ 7.71 & 1692.76 $\pm$ 4.79 & 1704.16 $\pm$ 7.89 \\
\hline
cRBM-1 & 1699.74 $\pm$ 11.60 & 1697.04 $\pm$ 17.24 & 1701.32 $\pm$ 8.32 & 1700.39 $\pm$ 10.00 & 1704.06 $\pm$ 5.28 \\
\hline
cRBM-3 & 1713.13 $\pm$ 8.78 & 1693.07 $\pm$ 5.39 & 1700.33 $\pm$ 6.77 & 1686.20 $\pm$ 10.91 & 1709.14 $\pm$ 8.70 \\
\hline
FC & 1700.01 $\pm$ 7.50 & 1697.11 $\pm$ 8.84 & 1702.02 $\pm$ 6.02 & 1702.20 $\pm$ 14.54 & 1702.95 $\pm$ 8.13 \\
\hline
\end{tabular}
\caption{\small An ablation study { showing the cut values} for different optimization algorithms and model architectures, tested on the graph instance with 150 nodes ($\text{UBD}=1784.89$) and batch size 1024. Stochastic natural gradient descent consistently outperformed all other optimizers, across all architectures considered.}
\label{tbl:optimization_algorithm_and_architecture}
\end{table*}

The role of optimization algorithm and model architecture was also investigated, focusing on the following optimizers (with and without natural gradient updates): Adadelta~\citep{zeiler-neurips12}, Adamax ($\alpha=5\times10^{-3}$)~\citep{kingma-iclr15}, Momentum ($\eta=5\times 10^{-2}$)~\citep{sutskever-pmlr13}, RMSprop ($\eta=5\times10^{-3}$),  SGD ($\eta=5\times10^{-2}$).
The architecture was chosen to be the restricted Boltzmann form \eqref{e:rbm} with hidden unit density $\alpha=m/n$ (RBM-$\alpha$), with real-valued weights (rRBM) and complex weights (cRBM).

The natural gradient descent \citep{amari1998natural, sorella_aps98} proved essential for converging to a good local optimum; results on cRBM-1 suggest that optimizers equipped with natural gradient updates consistently outperformed those without. The use of complex RBM yielded some improvement relative to the real-valued case, although the performance gap largely disappeared when natural gradient updates were applied. In addition, we found that architectures other than RBM can achieve high performance: the single-layer perceptron (FC) is compatible with RBM across all optimizers. In general, Stochastic Natural Gradient Descent consistently achieves optimal performance over all architectures, in comparison with other optimizers.

\section{Conclusions}
The Max-Cut approximation ratio achieved by natural evolution strategies is competitive with { widely used} solvers, although this comes at the expense of significantly increased computation time.  It will be interesting to investigate other combinatorial optimization problems, particularly from the spin-glass literature, which do not admit semidefinite program relaxations.

It is legitimate to inquire about possible advantages for classical stochastic optimization, given access to efficiently simulable subsets of quantum states, such as the complex Boltzmann machine. In the case of natural evolution strategies our ablative study indicates that the use of natural gradient descent levels the playing field between the quantum and classical variants.

\section{Acknowledgements}
Authors gratefully acknowledge support from NSF under grant DMS-2038030.

\appendix
\section{Supplementary material}
The following example illustrates how the complex restricted Boltzmann machine can efficiently represent probability distributions that are hard to represent using classical neural networks.

  Let $X=\{0,1\}^n$ be the set of all bitstrings of length $n$ and denote $P_n \subseteq X$ the subset consisting of all  $2^{n-1}$ bitstring with even Hamming weight. Define $p_n \in \mathcal{P}(X)$ as follows,
 \begin{equation}
     p_n = \frac{1}{|P_n|} \mathbbm{1}_{P_n} \enspace ,
 \end{equation}
where $\mathbbm{1}_{P_n} : X \to \{0,1\}$ is the indicator function for the subset $P_n$. Approximate representation of $p_n$ by a real-valued restricted Boltzmann machine requires a number $m$ of hidden units  scaling exponentially in $n$ \citep{montufar2011expressive} (see also \citep{stokes2019probabilistic}). In contrast, $p_n$ can be exactly represented by a complex RBM with $m=1$ hidden unit.

\bibliography{references}

\end{document}